\documentclass{nature}
\usepackage{siunitx}
\usepackage{graphicx}
\usepackage{textcomp}
\usepackage{lineno}
\usepackage[version=3]{mhchem}
\DeclareSIUnit\Molar{\textsc{m}}
\DeclareSIUnit\sccm{sccm}
\title{Mapping brain activity with flexible graphene micro-transistors}

\author{Benno M. Blaschke$^{1}$, N\'{u}ria Tort-Colet$^2$, Anton Guimer\`{a}-Brunet$^{3,4}$, Julia Weinert$^2$, Lionel Rousseau$^5$, Axel Heimann$^6$, Simon Drieschner$^1$, Oliver Kempski$^6$, Rosa Villa$^{3,4}$, Maria V. Sanchez-Vives$^{2,7}$ \& Jose A. Garrido$^{7,8}$*}

\begin{document}

\maketitle

\begin{affiliations}
 \item Walter Schottky Institut und Physik Department, Technische Universit\"at M\"unchen, Am Coulombwall 4, 85748 Garching, Germany

\item Institut d'Investigacions Biom\`{e}diques August Pi i Sunyer (IDIBAPS), 08036 Barcelona, Spain

\item Instituto de Microelectronica de Barcelona (IMB-CNM), CSIC, Campus UAB, 08913 Bellaterra, Barcelona, Spain
\item Centro de Investigacion Biom\`{e}dica en Red, Biomateriales y Nanomedicina (CIBER-BBN), Spain
\item ESIEE-Paris, ESYCOM, University Paris EST, Cit\'{e} Descartes BP99, 93160 Noisy-Le-Grand, France

\item Inst. Neurosurgical Pathophysiology, University Medical Center Mainz, Johannes-Gutenberg Univ. Mainz, Germany

\item ICREA, Pg. Llu\'{i}s Companys 23, 08010 Barcelona, Spain

\item Catalan Institute of Nanoscience and Nanotechnology (ICN2), CSIC and The Barcelona Institute of Science and Technology, Campus UAB, Bellaterra, 08193 Barcelona, Spain
\end{affiliations}
\noindent
\** joseantonio.garrido@icn2.cat

\newpage

\begin{abstract}

Establishing a reliable communication interface between the brain and electronic devices is of paramount importance for exploiting the full potential of neural prostheses\cite{Mehring.2003,Hochberg.2006, Kellis.2010, Scherberger.2005}. Current microelectrode technologies for recording electrical activity, however, evidence important shortcomings, e.g. challenging high density integration. Solution-gated field-effect transistors (SGFETs), on the other hand, could overcome these shortcomings if a suitable transistor material were available. Graphene is particularly attractive due to its biocompatibility, chemical stability, flexibility, low intrinsic electronic noise and high charge carrier mobilities\cite{ Bendali.2013,Fabbro.2016,Wang.2008,Hess.2011,Kim.2009}. Here, we report on the use of an array of flexible graphene SGFETs for recording spontaneous slow waves, as well as visually evoked and also pre-epileptic activity in vivo in rats. The flexible array of graphene SGFETs allows mapping brain electrical activity with excellent signal-to-noise ratio (SNR), suggesting that this technology could lay the foundation for a future generation of in vivo recording implants. 
\end{abstract}

Recording brain activity with high fidelity and decoding the enclosed information could enable the development of a new generation of neuroprosthetic devices for control of artificial limbs and motor rehabilitation, as well as brain-machine interfaces for communication and speech prostheses\cite{Mehring.2003,Lebedev.2006,Guenther.2009}. A major challenge is still the need of high-density, small recording sites that provide high spatial resolution with adequate signal-to-noise ratio (SNR) recordings to obtain high fidelity data for decoding as much information as possible. The most extended technology for in vivo recordings today uses microelectrode arrays (MEAs), mainly based on metals such as Pt and PtIr\cite{Obien.2014}. However, using MEAs for high-density recordings presents important drawbacks. Since the electrode impedance and noise are inversely proportional to the electrode size, a trade-off between spatial resolution and SNR has to be made. In addition, the high impedance of small electrodes creates enormous challenges in terms of on-chip multiplexing and, thus, for recording large numbers of electrodes in parallel. Further, the very small voltages of the recorded signals are highly susceptible to noise in the standard electrode configuration. For this reason, preamplification is required directly or very close to the electrode site. To overcome some of these issues, the electrode can be directly connected to the gate of an underlying transistor that converts the recorded voltage to current. This method facilitates multiplexing and provides a first amplification stage, which has been applied to demonstrate recordings from high-density flexible electrodes in in vivo experiments\cite{Viventi.2011}. However, the fabrication complexity is significantly increased and the additional electrical components required for the voltage-to-current conversion limit the integration density\cite{Viventi.2011}. 
Differently to the electrode recording configuration, Fromherz et al. showed that metal-oxide-semiconductor field-effect transistors (MOSFETs) where the gate metal is replaced with an electrolyte and an electrode, referred to as solution-gated field-effect transistors (SGFETs) or electrolyte-gated field-effect transistors, can be exposed directly to neurons and be used to record action potentials with high fidelity\cite{Fromherz.1991}. An important benefit of this recording configuration is the transistor's intrinsic signal amplification, which reduces the sensitivity to external noise. Further, the low impedance characteristic of the transistor configuration depends on the transistor geometry relations (width and length) and not the area (as is the case for the electrode configuration). This facilitates the implementation of multiplexing while allowing for down-scaling of the recording sites and maintaining low fabrication complexity.
To use transistors in long-term in vivo or chronic applications, several requirements have to be fulfilled by the substrate and the recording material: flexibility to avoid scar tissue formation, biocompatibility to avoid inflammation and toxicity, and stability in biological environments. In addition, the transistor's active material should also provide high transconductance, mainly governed by the charge carrier mobility and the capacitance of the transistor-electrolyte interface\cite{Hess.2011}, and a low electronic noise. These two parameters determine the minimum signal that can be detected and the SNR that can be achieved with this device. Besides silicon\cite{Fromherz.1991}, several other materials such as gallium nitride\cite{Steinhoff.2005}, diamond\cite{Dankerl.2009}, organic materials\cite{Benfenati.2013}, silicon nanowires\cite{Timko.2009} and more recently PEDOT:PSS\cite{Khodagholy.2013} and graphene\cite{Hess.2011b,Blaschke.2016}, have been tested for interfacing biological systems with transistors. However, gallium nitride, diamond and silicon introduce enormous challenges with respect to the integration with flexible substrates. Organic materials, on the contrary, can be integrated without major problems in flexible technologies but most of them have rather low charge carrier mobilities and are therefore not suitable for recordings with a high SNR. Interestingly, PEDOT:PSS electrochemical transistors, which use a sensing mechanism different from field-effect transistors, have been used to demonstrate in vivo recordings and are attracting significant interest despite the low matureness of this technology.
Graphene, on the other hand, should be ideally suitable for bioelectronic applications, due to its biocompability and chemical stability \cite{ Bendali.2013,Fabbro.2016,Wang.2008}. As a two dimensional material, the integration on flexible substrates is also unproblematic\cite{ Kim.2009, Kim.2010b}. The semiconductor-compatible fabrication of graphene SGFETs, together with the advances in the production of large-scale, high-quality chemical vapor deposition (CVD) graphene, makes the fabrication of large high-density graphene SGFET arrays for neural recordings possible\cite{Wu.2015}. The high charge carrier mobilities in graphene (typically well above 1000 cm²/Vs for CVD graphene) and the large interfacial capacitance of the graphene/electrolyte interface ($>$2\,\textmu F/cm\textsuperscript{2}) give rise to high transconductances, of more than 7\,mS/V\cite{Dankerl.2010,Hess.2011}. Together with a relatively low noise, this enables in vitro recordings with an excellent SNR\cite{Blaschke.2016}. While such values can also be obtained with PEDOT:PSS electrochemical transistors, graphene additionally exhibits a high transparency from the ultraviolet to the infrared, a key requirement for combining electrical measurements with optogenetic experiments\cite{Bonaccorso.2010, Park.2014,Kuzum.2014}. Recent publications also showed that graphene SGFETs are stable in cell culture environments\cite{Blaschke.2016}. Although the potential of graphene-based SGFET technology has been suggested in in vitro studies, so far no in vivo confirmation has been demonstrated. 
Here we present the fabrication of flexible arrays of graphene SGFETs and demonstrate in vivo mapping of spontaneous slow waves, as well as visually evoked and pre-epileptic activity in the rat.

Arrays of 16 SGFETs (transistor active area of W=20\,\textmu m, L=15\,\textmu m) were fabricated on polyimide substrate (figure 1(a)). A detailed description of the fabrication process can be found in the Methods. In vivo local field potential (LFP) measurements were performed in the brain of anaesthetized rats (figure 1(b)). After performing a craniotomy, the transistor array was placed on the surface of the rat visual cortex next to a 32-channel Pt MEA device (see figure 1b for an optical image of the arrangement). The transistors were characterized in vivo by measuring the drain-source current $I_{DS}$ as a function of the gate voltage $U_{GS}$ with fixed drain-source voltage. The transistor curves (figure 1c) exhibit the expected ambipolar V-shape of graphene transistors. From the transistor curve, the transconductance $g_m$ can be extracted (figure 1(c)). Defined as the derivative of $I_{DS}$ with respect to $U_{GS}$, $g_m$ describes the change in $I_{DS}$ induced by a small variation of $U_{GS}$. Consequently, the higher $g_m$, the larger the current change caused by a fluctuation of the electrical potential in the brain tissue next to the transistor. The detection limit and SNR of such potential fluctuations are determined by the transconductance and the intrinsic electronic noise of the transistors. The power spectral density of graphene SGFETs typically exhibits 1/f noise in the low frequency regime\cite{Blaschke.2016}. In order to estimate the SNR, the root mean square (rms) gate noise $U_{rms}$ is the most useful parameter. $U_{rms}$ was calculated as the standard deviation (STD) of the filtered transistor current in the case of no brain activity and then converted to a voltage using the transconductance. In post mortem recordings, values as low as 16\,\textmu V were measured for graphene micro-transistors. 

For the first neuronal recordings, pre-epileptic activity in the rat brain was induced by the local application of bicuculline\cite{DeFazio.2005}. Figure 2(a) shows an exemplary simultaneous recording of the pre-epileptic activity using a transistor (red), a 50\,\textmu m (blue) and a 10\,\textmu m (black) diameter Pt electrode. All three devices recorded interictal spikes that coincided temporally. The graphene transistor (active area 300\,\textmu m\textsuperscript{2}) and the large Pt electrode (active area 1962\,\textmu m\textsuperscript{2}) recorded significantly larger peaks than the small Pt electrodes (active area 78\,\textmu m\textsuperscript{2}). In figure 2(b) a single spike recorded by a transistor with a time-frequency analysis in the background is shown depicting the increased power at low frequencies during bicuculline-induced activity. It is worth noticing that the graphene SGFETs were operated with zero gate bias, thus no voltage had to be applied between the transistor and the brain; this is in contrast to transistors based on PEDOT:PSS, in which a gate voltage is necessary to bias the transistors in the operating conditions\cite{Khodagholy.2014}.

To compare the ability of the different devices to detect such pre-epileptic activity against the background brain activity, the SNR was calculated for every transistor and electrode as described in the Methods. SNR values of up to 72 with an average of $62\pm5.8$ were estimated for five graphene SGFETs. A maximum SNR of 34 with an average of $26\pm5.5$ for 8 small Pt electrodes and a maximum value of 75 with an average $53\pm11$ for 13 large Pt electrodes were obtained. The obtained SNR values for the bicuculline-induced activity show that the graphene SGFETs can compete with-state-of-the-art Pt electrodes of both sizes. The small voltages recorded by the electrodes are very susceptible to noise. Therefore, they have to be pre-amplified as close to the recording site as possible, typically this is done at the connector of the electrode. In contrast, the transistors were connected to the amplification setup by a 30 cm long unshielded wire without showing problems from externally coupled noise, evidencing the advantage of the intrinsic signal amplification of the transistor concept. The SNR of the graphene FETs is similar to that of graphene electrodes used in vivo; however, these graphene electrodes were significantly larger and thus their SNR performance will decrease when downscaled to the graphene transistor size\cite{Kuzum.2014}. Pt-based high density electrode arrays with on-chip multiplexing achieve similar noise values, however with an area that is two orders of magnitude larger than our graphene FETs\cite{Viventi.2011}.

The current STD background of the graphene transistor was similar (around 5\,nA) to PEDOT:PSS transistors, though the recorded potentials in the GAERS rat model used to test the PEDOT:PSS transistors were significantly larger resulting in a higher SNR\cite{ Khodagholy.2013}. 
 
To demonstrate the mapping capability of the SGFET array, figure 2(c) shows a map of the averaged interictal spikes (see Methods for averaging procedure) of electrodes (50\,\textmu m blue, 10\,\textmu m black) and graphene transistors (red). The transistors mainly show homogeneous peaks, with an amplitude similar to the 50\,\textmu m electrodes and a slightly higher SNR ratio. The bicuculline-spikes recorded by the 10\,\textmu m electrodes are significantly smaller and also have a significantly lower SNR. To explore the ambipolar behavior of the graphene transistors, the graphene SGFETs were biased in different regimes. figure 2(d) shows a typical transistor curve superimposed to the averaged recorded current of bicuculline-induced interictal spikes in the different bias regimes. In the hole conduction regime of the graphene transistor ($U_{GS}=$150\,mV), the transconductance is negative and the negative voltage during the epileptiform discharges results in a positive signal in the transistor current ($I_{DS}=U_G\cdot g_m$). In contrast, $g_m$ is positive in the regime of electron conduction ($U_{GS}=$600\,mV) and the negative voltage peak results in a negative current peak. In the vicinity of the Dirac point, where the transconductance is close to zero, the recorded activity is almost zero. This ambipolar behavior can be very useful to distinguish between biological signals and external noise that is coupled into the measurement system. In addition, the possibility to bias the transistor offers a way to tune the device response in order to maximize the recorded signal, which is not available in the case of electrodes.

In order to probe our recording system with more physiological and smaller amplitude signals, we recorded two types of neural activity from the primary visual cortex: spontaneous slow oscillations typical from slow-wave sleep and deep anesthesia and visually evoked responses. Under these conditions, the spontaneous cortical activity is characterized by a slow (\textless 1Hz) alternation between active and silent states\cite{Steriade.1993, SanchezVives.2014, Bettinardi.2015}. Figure 3(a) shows traces of LFP simultaneously recorded during spontaneous activity for each device type. In this case, the SNR was defined as the amplitude of the slow-wave divided by the standard deviation in the silent periods between waves. The graphene transistors and the 50\,\textmu m Pt electrodes show average SNR values of $9.85\pm0.67$ and $8.33\pm1.05$, respectively, whereas the smaller Pt electrodes only exhibit a SNR of a $6.02\pm0.68$. 

In a following experiment, the recording of a visually evoked response was studied by inducing light stimulation using a light-emitting diode (LED), as described in the Methods. Since the recording arrays were placed on the visual cortex, the light stimulation induced a visually evoked response that can be detected by the recording devices. Figure 3(b) shows the averaged response (red), calculated as described in the Methods, and a single response (light red) recorded by a graphene SGFET and the single response of a Pt electrode (light blue) and the averaged response (blue). Approximately 30\,ms after the light onset, a steep increase is observed followed by a slower decay, as reported previously\cite{Wang.2014}. While the averaged signals provide an excellent SNR, even the single recordings can be used to clearly identify the evoked response. The map in figure 3(c) shows the averaged response of several electrodes and transistors. The significant variation in the amplitudes is caused by the local nature of the evoked brain activity. Therefore, the data does not allow a proper comparison in terms of SNR. 

To investigate the biocompatibility of graphene implants we performed an immunohistology study using samples with graphene on polyimide and only polyimide that were implanted subdurrally in rats (see Methods). The upper panel of figure 3(d( shows a typical microscope image of subdural brain tissues immunostained for microglial and astroglial markers (Iba-1, GFAP) 28 days after implantation of a graphene implant. The immunohistology study evaluated microglial activation as a sign of inflammatory processes by quantifying morphological changes of microglia based on two indices: solidity and circularity (see Methods). For comparison, sham-operated animals were sacrificed 4 days after surgery to show acute surgical trauma effects. When compared to naive rats without surgical trauma both circularity and solidity showed statistically significant inflammatory reactions in these animals, which however were minor. E.g. circularity increased from 0.039 to 0.063 (when considering the possible maximum close to 1.0) as shown in figure 3(d). Graphene implants did not show any significant changes of circularity or solidity at any of the time points tested as compared to naive rats or polyimide samples without graphene confirming the biocompatibility of the graphene devices.

In summary, we demonstrated the successful recording of in vivo brain activity using flexible arrays of graphene-based SGFETs. Recording LFP during spontaneous slow oscillations, visually evoked activity, and pharmacologically-induced pre-epileptic spikes, our results show that graphene transistors can compete with existing state-of-the-art microelectrode-based recording technologies, while additionally offering advantages such as intrinsic signal amplification and the possibility for down-scaling and high-density integration. High-density recordings of brain activity over large areas is an important challenge that has to be overcome in order to enable the development of a new generation of neuroprosthetic devices. The results in this work demonstrate that technologies based on flexible graphene field-effect transistors are uniquely positioned to offer such high-density recordings when combined with already demonstrated wafer-scale very-large-scale integration (VLSI) compatible fabrication of graphene transistors and advanced on chip multiplexing\cite{Rahimi.2014,Viventi.2011}. Together with functionalized graphene transistors for the detection of neurotransmitters\cite{Hess.2014} or optogenetics, these technologies could provide deeper insights into biological processes. The combination of graphene with other 2D materials, e.g. boron nitride substrates, is expected to further enhance the mobility and decrease the noise of flexible SGFETs resulting in an improved SNR performance\cite{Dean.2010, Kayyalha.2015} possibly enabling the detection of single unit activity from the brain surface. Future experiments should aim at the combination of graphene SGFET recording sites with on chip multiplexing based on other 2D materials or CMOS technology.
Based on the biocompatibility of graphene implants and taking into account the large room for improving the performance of graphene-based flexible field-effect transistors, for instance by improving processing technology and material quality, we foresee a very rapid advance of graphene technologies in neural functional interfaces.

\begin{methods}
\subsection{Transistor fabrication and electrode design}

A sacrificial 500\,nm aluminum layer was sputtered on a 4-inch silicon wafer. Afterwards, a 7 to 10\,\textmu m thick biocompatible\cite{Lago.2005,Klinge.2001, Yeager.2008} polyimide 2611 layer (HD MicroSystems) was spin-coated and cured under nitrogen atmosphere at 350\,$^\circ$C. Titanium tungsten (20\,nm) and gold (100\,nm) was deposited by sputtering and then structured by optical lithography and etching to form drain and source contacts. CVD graphene was transferred to the wafer using a PMMA wet etching process as reported previously\cite{Blaschke.2016}. After photolithography, the graphene was structured using an oxygen plasma in a reactive ion etching system. A second metal layer of 900\,nm gold was sputtered on the sample and structured by photolithography and etching. A less than 2\,\textmu m thick SU8 resist was spin-coated on the sample and structured by optical lithography to create openings defining the active transistor area. To ensure complete insulation, a part of the graphene next to the contacts is also covered with 2\,\textmu m SU8, giving rise to an access resistance caused by the underlying ungated graphene. In order to define the shape of the implants, a 500\,nm-thick aluminium layer was sputtered on the wafer and structured by photolithography and reactive ion etching (\ce{Cl_2}, \ce{BCl_{3}} and \ce{N_{2}}). After defining the shape of the implants by another reactive ion etching step using \ce{O_2} and \ce{N_2}\textbf{•}, the aluminum was etched away. The samples were released by electro-erosion of the sacrificial aluminum layer. The samples were bonded with a two component conductive epoxy glue to a custom-designed PCB and connected with wires to the measurement setup.
For in vivo recordings 32-channel arrays of platinum electrodes (fabricated by CNR-IMM, Rome, Italy) were used. The diameter of the recording site of the platinum electrodes was either 10\,\textmu m (8 channels) or 50\,\textmu m (24 channels). The microelectrodes were fabricated by embedding a tri-layer Cr/Au/Pt (200\,nm thick) into polyimide HD2611 (HD MicroSystems) layers, reaching a final thickness of 8\,\textmu mm. See Castagnola et al. for details \cite{Castagnola.}.

\subsection{Experimental procedures for in vivo measurements}
For the in vivo experiments, adult male Wistar rats were placed in an anesthesia induction chamber for 5 minutes at 100\% of \ce{O_2}. Next, anesthesia was induced by raising the isoflurane concentration to 5\% (0.6\,L/min, 1\,bar) for 5 more minutes always watching out respiration. We next set the concentration of isoflurane to 3\% for one more minute before the rat was placed in the stereotaxic apparatus with a mask delivering a mixture of isoflurane and oxygen. For the rest of the surgery, 3\% of isoflurane was used to maintain deep anesthesia. A subcutaneous injection of atropine (0.05\,mg/kg) was given to prevent respiratory secretions. Methylprednisolone (10\,mg/kg) was injected (i.p.) to prevent inflammation. Rectal temperature was maintained at $37^\circ$C. A craniotomy was performed to access the primary visual (V1) cortex (7.3\,mm AP, 3.5\,mm ML) of the left hemisphere\cite{ Paxinos.2007}. The graphene transistor array and the 32-channel Pt MEA were placed on the cortex. To evoke visual responses in the cortex, a light-emitting diode (LED) was placed in front of the right eye (contralateral to the recording site) of the rat and a flash of 100\,ms was automatically delivered every 4 to 5 seconds. In some recordings, \SI{200}{\micro\Molar} bicuculline methiodide (Sigma), a \ce{GABA_a} receptor blocker that is broadly used to pharmacologically reduce inhibition in the brain and thus generate epileptiform activity, was directly applied to the surface of the cortex. For more detailed methodology of the in vivo experiments see\cite{Reig.2007,Amigo.2015,Reig.2015}. Experiments on four animals were performed, the presented data are all from the same animal.
All experiments were supervised and approved by the University Committee and were carried out in accordance with the present laws of animal care, EU guidelines on protection of vertebrates used for experimentation (Strasbourg 3/18/1986) and the local law of animal care established by the Generalitat of Catalonia (Decree 214/97, 20 July).

\subsection{Data acquisition}
A custom-built setup was used for transistor characterization and neural recordings with the transistor array. In a first step, the transistor current is transformed to a voltage and low-pass filtered at 15\,kHz using an operational amplifier feedback loop. For the neural recordings an additional amplification by a factor of 100 and high-pass filtering at 2.4\,Hz is performed. The signal is then recorded by a National Instruments LabVIEW DAQ Card and a LabVIEW program. For the device characterization the drain-source and gate voltage were applied by the DAQCard, and for the in vitro and in vivo recordings batteries were used to reduce the electronic noise. All measurements were performed in a Faraday cage. The gate voltage was applied to a Ag/AgCl reference electrode. In case of simultaneous electrode and transistor recordings, the gate voltage was set to zero. For non-zero gate voltages the MEA was disconnected. 
In the in vivo electrode recordings the signal from the 32 electrodes were amplified with a multichannel system using a MPA8 miniature preamplifier (Multi Channel Systems, input impedance $10^{12}\,\Omega$) and digitized at 10\,KHz with a CED 1401 POWER3 (Cambridge Electronic Design) acquisition board and Spike 2 software.

\subsection{Data treatment}
Data filtering and analysis were performed with MATLAB. All data were low-pass filtered at 200\,Hz using a first-order Butterworth filter. In addition, digital notch filters were used to remove 50\,Hz noise and its overtones. For the spontaneous slow oscillation recording, the electrode data were additionally high-pass filtered using a first-order Butterworth filter with a cut-off frequency of 2.4\,Hz to allow comparison with the transistor signals.

For the estimation of the SNR ratio, the bicuculline-induced peaks were automatically detected, and the peak-to-peak amplitude was extracted. The standard deviation $U_{STD}$ was calculated during the non-spike periods (averaged across non-spike periods after windowing each non-spike period in small non-overlapping windows of 50\,ms). Slow-wave activity occurring during the inter-spike periods was discarded for the STD computation. For each peak, the SNR given by $SNR=A_{peak-to-peak}/U_{STD}$ was calculated and the SNRs were averaged afterwards. For the SNR estimation of the spontaneous slow oscillations, the peak-to-peak amplitude of the slow-wave was extracted and divided by the standard deviation of the signal during the silent periods between waves. Time-frequency analysis was performed by continuous wavelet transform with a Paul wavelet using MATLAB's wavelet toolbox.

\subsection{Procedures for the histology study}

Samples were fabricated on 4-inch silicon wafers. Polyimide deposition followed by graphene transfer was performed as for the graphene transistors. Two circular sheets of graphene (1\,mm diameter, separated by 200\,\textmu m) were defined by optical lithography and oxygen plasma. The definition of the implant shape and release were done as for the transistor devices. Presence of graphene on the samples was verified by conductance measurements using a tip probe station and fluorescence microscopy. Samples were implanted subdurally in Wistar rats using standard microsurgical techniques. Implantations were performed as described by Henle et al. 2011 under general anesthesia (Medetomidin, Ketamine, and Tramadol for intra- and post- operatively analgesia)\cite{Henle.2011}. The surgical technique was slightly modified with the dura-mater only incised with microscissors to slide in the electrodes beneath the dura. The bone flap was reinserted and fixed with tissue glue (Histoacryl). Rats were sacrificed after 14, 28 or 84 days. Biocompatibility was tested by immunohistology of subdural brain tissues for microglial and astroglial markers (Iba-1, GFAP). To do so, cryo-sections were allowed to defrost for 30 minutes at room temperature and rinsed with Triton-PBST. Fluorescence-staining for IBA-1 was performed with a microglia-specific antibody 'anti Iba1' (rabbit anti ionized Calcium binding Adapter Molecule 1, 1:100, Wako, USA) and Alexa Fluor 568\textsuperscript{\textregistered} donkey anti-rabbit (1:100, Life Technologies, Carlsbad, USA) as secondary antibody. Primary and secondary antibodies for specific glial-fibrillary acidic protein were mouse anti-glial-fibrillary acidic protein (GFAP, 1:100, BD Pharmingen, Becton Dickinson \& Comp., USA) and donkey-anti-mouse (1:20, Life Technologies, Carlsbad, USA). Primary antibodies were incubated 24 hours at 3$^\circ$C in the dark. After rinsing (3 times, 5 min with PBST) secondary antibodies remained 2h on the slices. Then cell nuclei were stained by 4´,6-diamidin-2-phenylindol (DAPI, 1:1000, Carl Roth, Germany). Microglial activation as a sign of inflammatory processes was quantified by evaluating morphological changes of microglia which changed from resting state with many branches to an activated state charaterized by loss of branches and rounding of the cell body. This was done by measuring cell perimeters and areas using ImageJ software. Cell perimeter decreases with activation. From these two parameters the 'circularity' can be calculated: 4*$\pi$*area/perimeter\textsuperscript{2} \cite{Kozlowski.2012} which increases with activation (1=maximum). Another index is 'solidity' (cell area/convex area) \cite{Soltys.2001}, which also increases with activation to a maximum of 1. All experiments were performed under animal welfare guidelines, and were approved by the local ethics committee (Landesuntersuchungsamt Koblenz, Germany, approval code: 23 177-07/G12-1-029).

\end{methods}

\newpage

\section{References}
\bibliography{bibliography}

\begin{addendum}
 \item We thank Guillermo Fortunato, Marco Marrani and Luca Maiolo from CNR-IMM in Rome, Italy for providing the platinum arrays. We acknowledge T. Wimmer and M. Prexl for help with preliminary experiments on flexible graphene SGFETs and F. Hrubesch for taking pictures. J.A.G., B.M.B., and S.D. acknowledge support by the German Research Foundation (DFG) in the framework of the Priority Program 1459 Graphene, the Nanosystems Initiative Munich (NIM), and the Graphene Flagship (Contract No. 604391). J.A.G., B.M.B., S.D., L.R., A.H and O.K. acknowledge support by the European Union under the NeuroCare FP7 project (Grant Agreement 280433). N.T.C., J.W., and M.V.S.-V. acknowledge support by the EU FF7 FET CORTICONIC contract 600806 to M.V.S.-V.. B.M.B., N.T.C., A.G.B., J.W., S.D., R.V., M.V.S.-V. and J.A.G. received funding for this project from the European Union's Horizon 2020 research and innovation programme under grant agreement No 696656. ICN2 acknowledges support from the Severo Ochoa Program (MINECO, Grant SEV-2013-0295).
\item[Author contributions] B.M.B., N.T.C, J.W., A.G.B, A.H., O.K., R.V., M.V.S.-V. and J.A.G planned and designed the experiments, B.M.B, L.R, A.G-B and S.D. fabricated the devices, B.M.B., N.T.C, J.W., A.G.B., M.V.S.-V. and J.A.G. performed the recording experiments and analysed the data, A.H. and O.K. performed the histology study, B.M.B., N.T.C, J.W., A.H., O.K., M.V.S.-V. and J.A.G. wrote the manuscript.
 \item[Competing Interests] The authors declare that they have no
competing financial interests.
 \item[Correspondence] Correspondence and requests for materials
should be addressed to J.A.G \linebreak (email: joseantoniogarrido@icn2.cat).
\end{addendum}

  \renewcommand{\baselinestretch}{1}\normalsize

\begin{figure}[h!t]
\centering
\includegraphics[width=\textwidth]{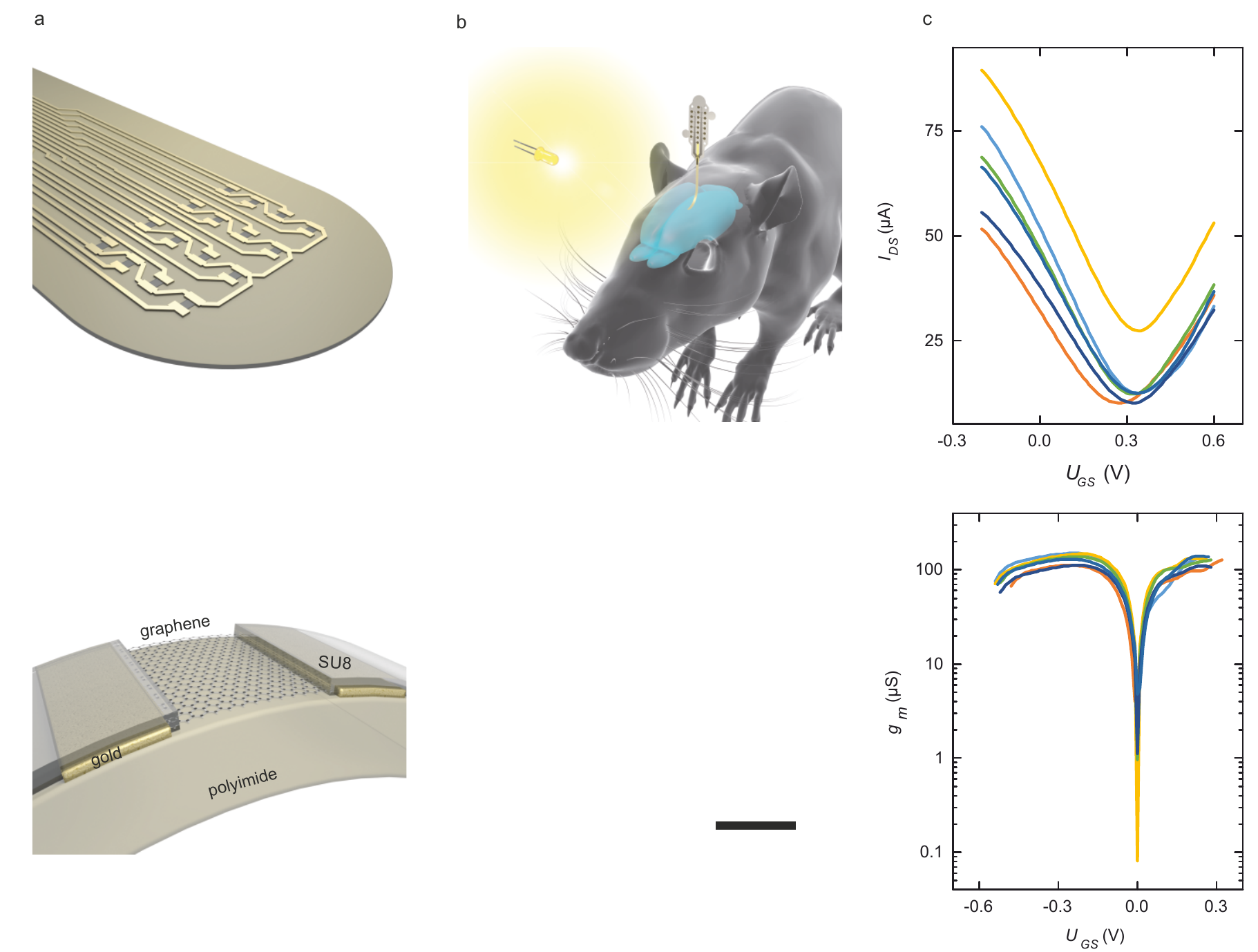}
\caption{\textbf{a)} Upper panel: Representation of the head of a graphene implant showing a 4x4 graphene transistor array and feed lines. Lower panel: Cross section of a graphene transistor with graphene between the source and drain contact that are covered by an insulating SU8 photoresist. \textbf{b)} Upper panel: Representation of the implant placed on the surface of the rat's brain. Lower panel: Microscope image of a MEA with Pt electrodes (a) and the graphene device (b) next to it. Scale bar is 1.25\,mm. \textbf{c)} In vivo characterization of devices. Upper panel: Transistor current $I_{DS}$ as a function of the gate voltage $U_ {GS}$ for a fixed drain-source voltage $U_{DS}=200mV$; different colors represent different transistors. Lower panel: Resulting transconductances.}
\end{figure}

\newpage

\begin{figure}[h!t]
\centering
\begin{minipage}[c]{0.66\textwidth}
\centering
\includegraphics[width=\textwidth]{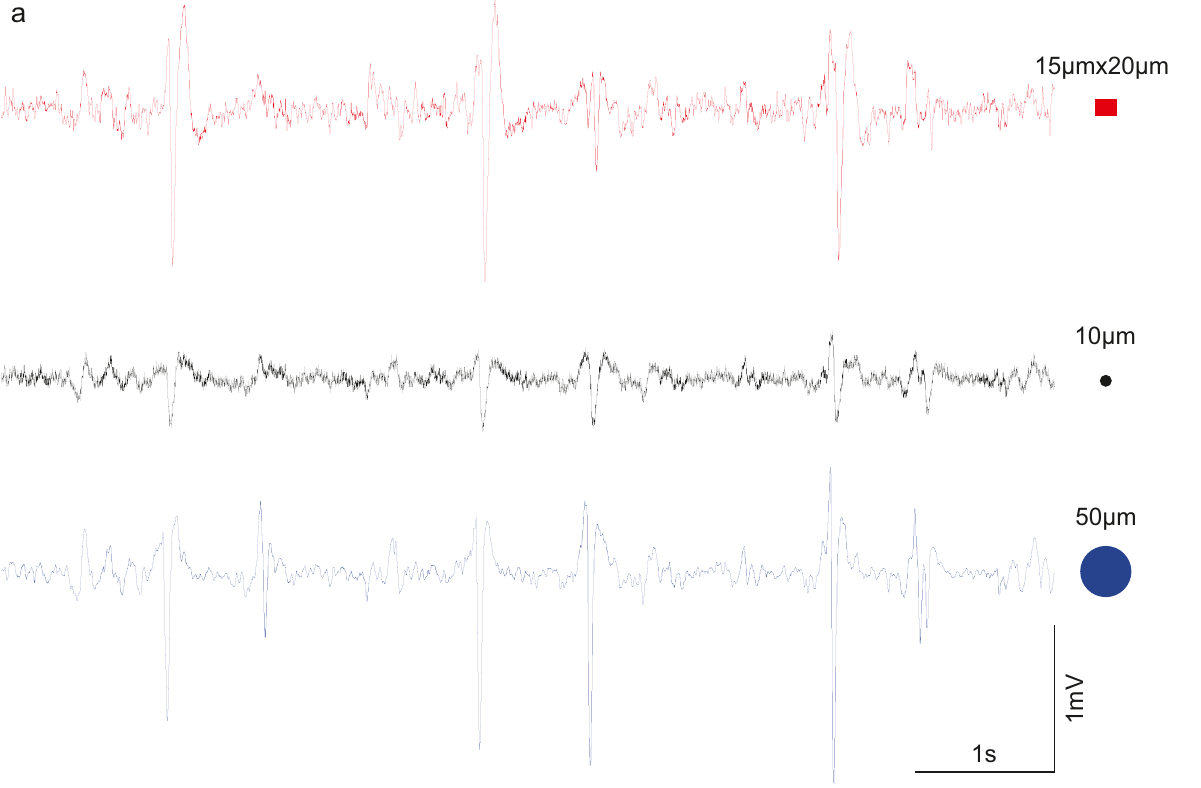}
\end{minipage}
\begin{minipage}[c]{0.33\textwidth}
\centering
\includegraphics[width=\textwidth]{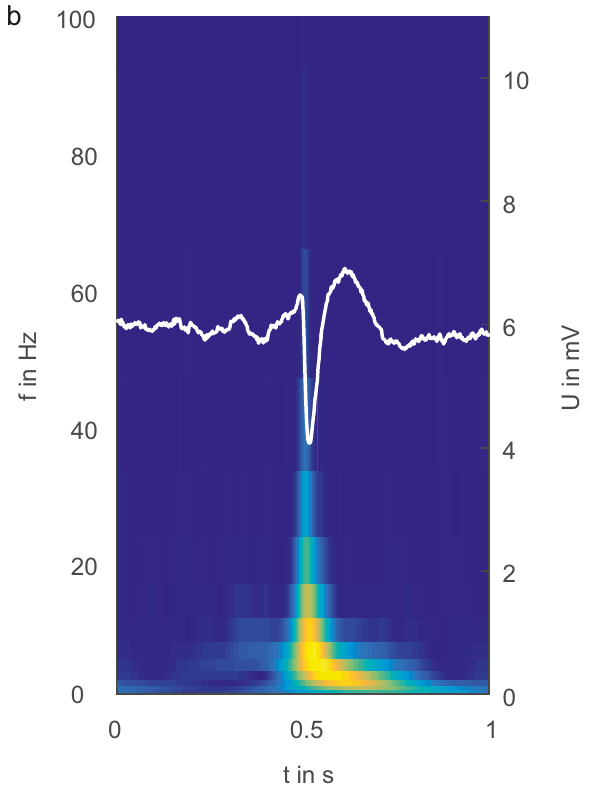}
\end{minipage}

\vspace{0.5cm}

\begin{minipage}[c]{0.66\textwidth}
\centering
\includegraphics[width=\textwidth]{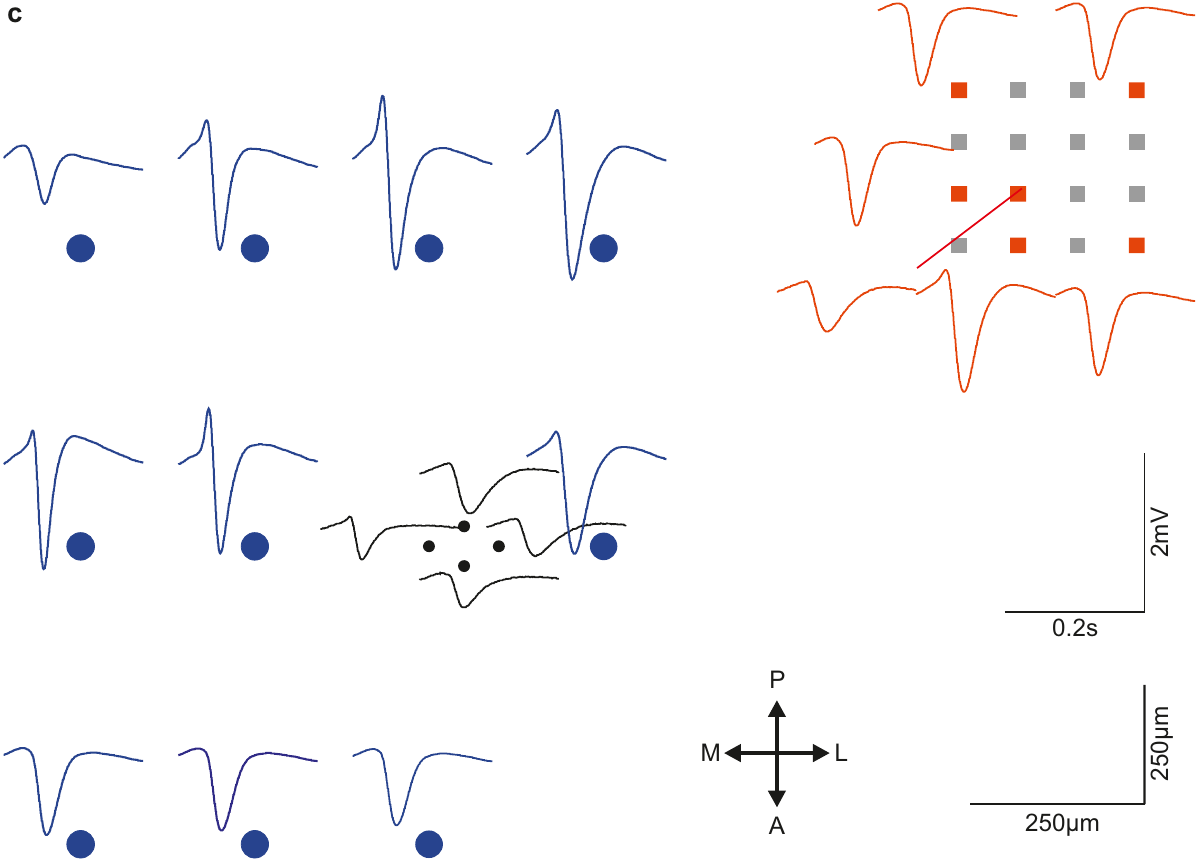}
\end{minipage}
\begin{minipage}[c]{0.33\textwidth}
\centering
\includegraphics[width=\textwidth]{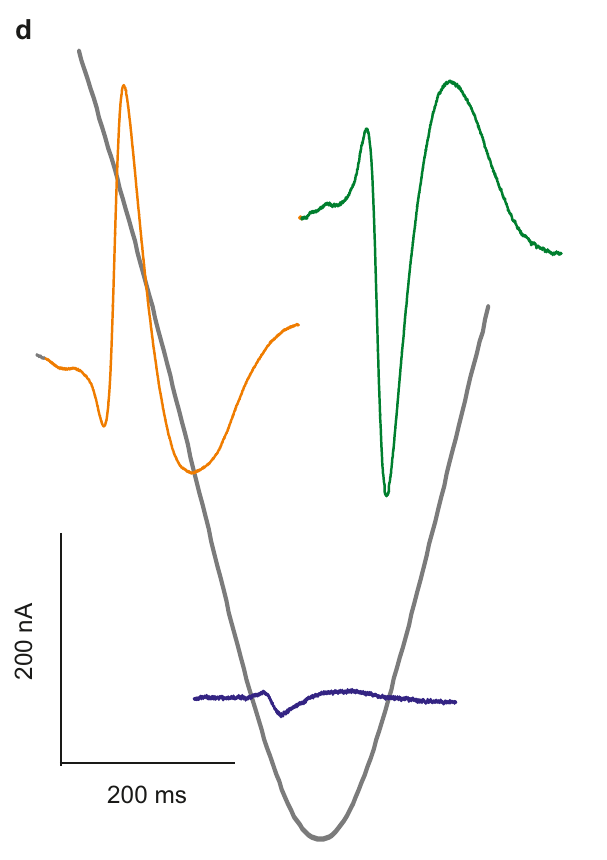}
\end{minipage}
\caption{\textbf{a)} Simultaneous recordings of a graphene transistor (red), a 10\,\textmu m (black) and a 50\,\textmu m (blue) diameter Pt electrodes showing bicuculline-induced brain activity. The transistors were biased with $U_{DS}=200$\,mV and the gate voltage was connected to the electrode ground. The shape and dimension of the recording site is shown for comparison. \textbf{b)} Single bicuculline-spike recorded by a graphene transistor with the time-frequency analysis in the background. \textbf{c)} Pre-epileptic discharges in bicuculline mapped onto the locations of electrodes and transistors. Arrows indicate anterior (A), posterior (P), lateral (L) and medial (M) directions on the cortical surface. \textbf{d)} A transistor curve together with averaged bicuculline-spike recorded in current by a graphene SGFET in the electron (orange) and hole (green) regime and in the vicinity of the Dirac point (purple).} 
\end{figure}

\newpage

\begin{figure}
\centering
\begin{minipage}[c]{0.66\textwidth}
\centering
\includegraphics[width=\textwidth]{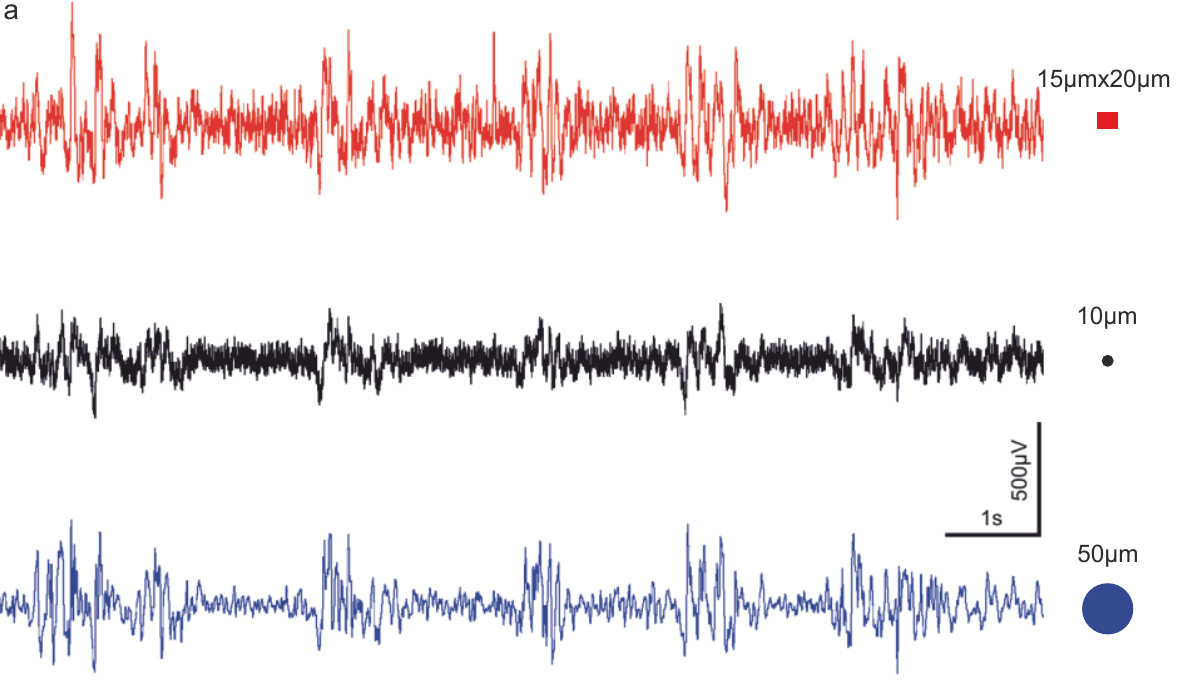}
\end{minipage}
\begin{minipage}[c]{0.33\textwidth}
\centering
\includegraphics[width=\textwidth]{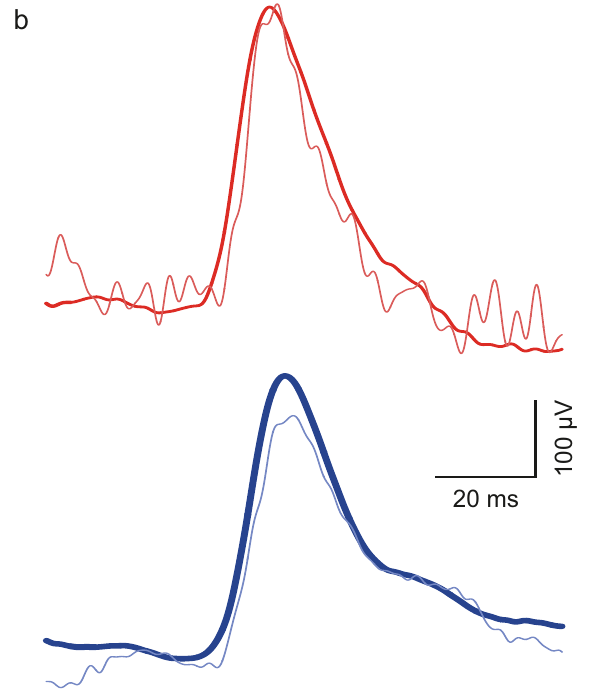}
\end{minipage}

\vspace{0.75cm}

\begin{minipage}[c]{0.66\textwidth}
\centering
\includegraphics[width=\textwidth]{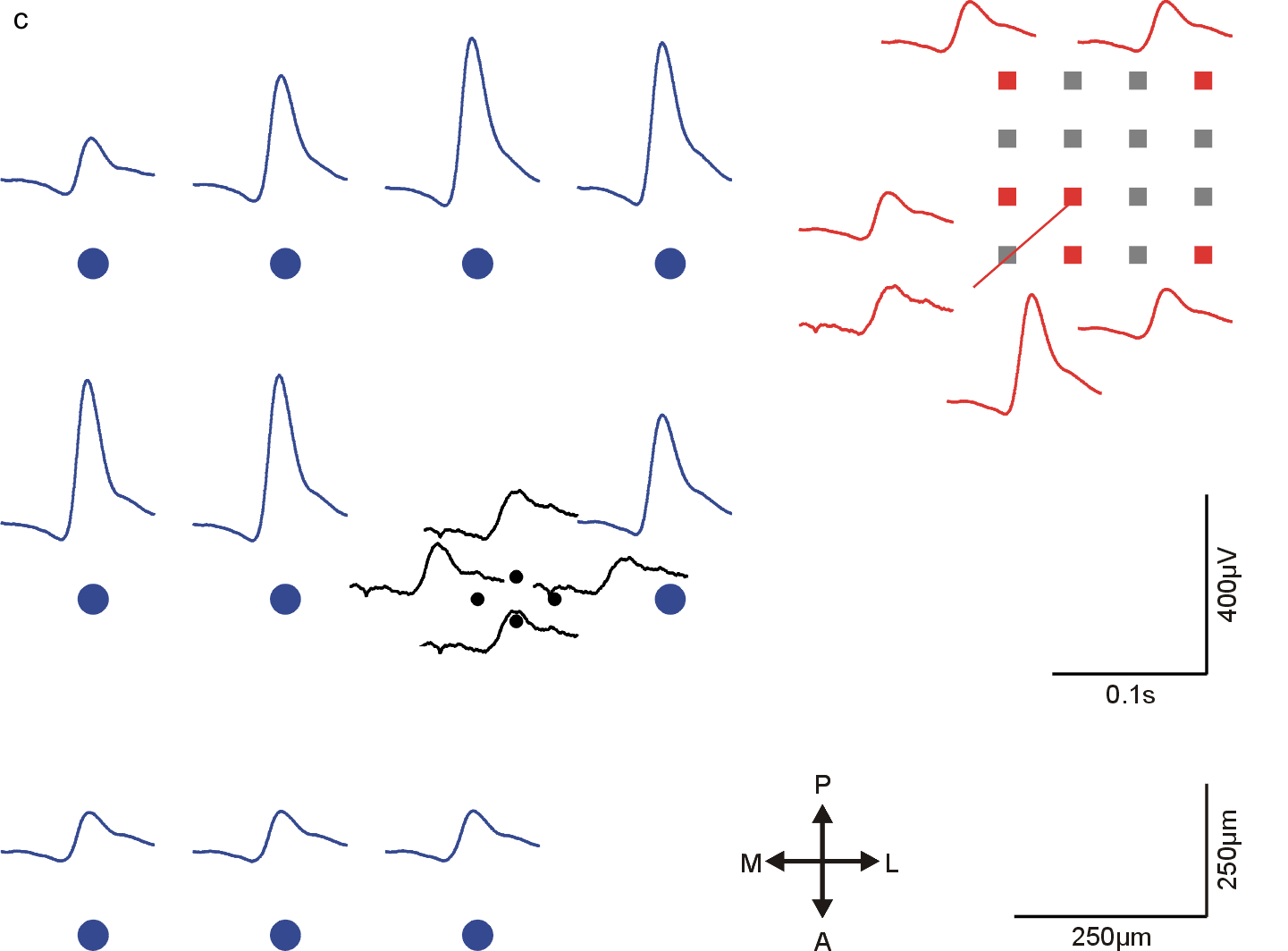}
\end{minipage}
\begin{minipage}[c]{0.33\textwidth}
\centering
\includegraphics[width=\textwidth]{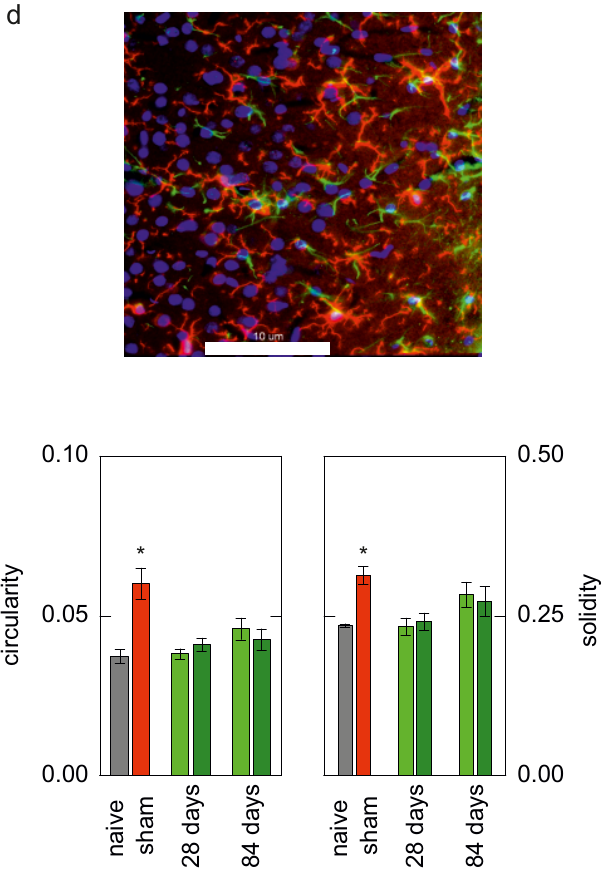}
\end{minipage}
\caption{\textbf{a)} Simultaneous recordings of spontaneous brain activity under deep anesthesia with a graphene SGFET (red), a 10\,\textmu m (black) and a 50\,\textmu m (blue) diameter Pt electrode showing slow oscillations of the LFP. \textbf{b)} Recording of a single event (light red) and averaged response (red) of 66 events recorded by a graphene SGFET induced by visual stimulation. Same below but obtained with a 50 with a 50\,\textmu m Pt electrode. \textbf{c)} Spatial map of the averaged visually evoked responses recorded by electrodes and transistors during visual stimulation with a light-emitting diode. Arrows indicate anterior (A), posterior (P), lateral (L) and medial (M) directions on the cortical surface. \textbf{d)} Upper panel: Typical microscope image of immunostained subdural rat brain tissue 2\,\textmu m below dura at the site of implantation 28 days after implantation of a graphene on polyimide sample. Colour code: Blue is DAPI nuclear stain, red is Iba-1 (microglia), and green is GFAP (astrocyte). Scale bar is 10\,\textmu m. Lower panel: Circularity and solidity (indices for inflammatory processes) for naive rats, 4 days after sham-operation and after 28 and 84 days after implantation of polyimide (light green) and graphene on polyimide implants (dark green). Results of increased inflammation after sham-operation are statistically significant (* t-test: p\textless 0.05 vs. naive animal.)}
\end{figure}

\end{document}